\documentclass[a4paper]{JHEP3} %

\usepackage{bbm,amsmath,graphicx,amssymb}
\usepackage{bm}
\usepackage{cite}
\usepackage{verbatim}
\usepackage{slashed}
\usepackage{epsf}

\def\spa#1.#2{\left\langle#1\,#2\right\rangle}
\def\spb#1.#2{\left[#1\,#2\right]}
\def\oneloop{{(1)}}

\def\nn{\nonumber}

\def\stamp{--- {\bf \today} --- {\bf \jobname.tex}}

\def\fs_#1{\mathfrak{s}(#1)}

\def\BE{\begin{equation}}
\def\EE{\end{equation}}
\def\spa#1.#2{\left\langle#1\,#2\right\rangle}
\def\spb#1.#2{\left[#1\,#2\right]}
\def\lor#1.#2{\left(#1\,#2\right)}

\newcommand\fverb{\setbox\fverbbox=\hbox\bgroup\verb}
\newcommand\fverbdo{\egroup\medskip\noindent%
			\fbox{\unhbox\fverbbox}\ }
\newcommand\fverbit{\egroup\item[\fbox{\unhbox\fverbbox}]}
\newbox\fverbbox

\title{Monodromy--like Relations for Finite Loop Amplitudes}

\author{ N.~E.~J. Bjerrum-Bohr$^a$, P. H. Damgaard$^a$, H. Johansson$^b$ and
{T. S{\o}ndergaard$^a$}\bigskip\\
{$^a$\small Niels Bohr International Academy and Discovery Center,\\
The Niels Bohr Institute, \\ Blegdamsvej 17, DK-2100 Copenhagen,\\
Denmark \\
{\tt email:}
\{bjbohr;phdamg;tsonderg\}@nbi.dk
\bigskip~\\$^b$\small Institut de Physique Th{\'e}orique, \\
CEA-Saclay,\\
F-91191 Gif-sur-Yvette,\\ France \\
{\tt email:} henrik.johansson@cea.fr}}

\received{\today} 		
\accepted{\today}		

\preprint{Saclay--IPhT--T11/041}

\abstract{We investigate the existence of relations for finite one-loop amplitudes in Yang-Mills theory. 
 Using a diagrammatic formalism and a remarkable connection between tree and loop level, we deduce sequences of amplitude relations for any number of external legs.
}


\begin{document}

\section{Introduction}
In the ongoing efforts to compute gauge theory and gravity scattering amplitudes, a consistent pattern is the emergence of hidden structures that imply a remarkable simplicity. New notions  and new formalism reduce the complexity of calculations from a variety of different directions. At tree-level, old and new computational techniques have almost trivialized the computation of helicity amplitudes for any multiplicity~\cite{BerendsGieleRecusion,BCFW,DrummondHennSolution}. Similarly, at the one-loop level highly-refined methods exists for determining amplitudes in massless theories~\cite{BCF,Forde,blackhat,NGluon}. The emergence of novel structures can have far-reaching consequences, potentially reshaping the way one organizes the theory beyond tree and one-loop level. Recent examples of this is the dual conformal symmetry of planar ${\cal N}=4$ super-Yang-Mills theory~\cite{DualConformalSymmetry}, and the conjectured duality between color and kinematical structures in gauge theories~\cite{BCJ, BCJloops, BCJother}.

While much effort has been devoted to the study of supersymmetric loop amplitudes, one class of non-supersymmetric loop amplitudes stands out as being very special. These are the  {\em finite} one-loop amplitudes of Yang-Mills theory in four dimensions, where the helicity constraints force the corresponding tree-level amplitudes to vanish.  The simplest such are the pure-gluon amplitudes  with helicity $(+++\cdots+)$ and $(-++\cdots+)$, which we will refer to as 'all-plus' and 'one-minus', respectively. The four-dimensional unitarity cuts vanish for these amplitudes, hence there can at most be poles. Logarithmic functions with branch cuts cannot be present, only rational terms contribute to the amplitudes. In fact, as can be expected, they resemble tree-level amplitudes in many ways~\cite{BernAllPlus,Mahlon}. This makes them natural objects to study in search of  structures and relations which may carry over from tree level. 

In this paper we explore several different types of relations for the finite one-loop gluon amplitudes of all-plus and one-minus helicity configurations. As part of the sequences of relations we discover, an interesting pattern emerges that links the observed one-minus relations with the intersection of the tree-level and all-plus relations. Furthermore there are indications that the totality of relations between amplitudes may be such that there exists a minimal basis of $(n-3)!$
amplitudes, thus mimicking what holds at tree level~\cite{BDVmonodromy}. This reduction of independent amplitudes can perhaps be deduced from string theory, similar to the tree level case. However, the detailed relations responsible for the observed reduction are far more complicated than at tree level, and typically the relations are helicity dependent. We will refer to the totality of relations we uncover as 'monodromy-like' because they include relations with both constant and momentum-dependent factors. This is in parallel with what occurs at tree-level, where string theory monodromy simultaneously explain both types of relations.

Our paper is organized as follows. In section~\ref{AmplPropsSection} we remind the reader of the basic properties and relations of both tree-level and one-loop color-ordered amplitudes. In section \ref{allplusSection} we focus on the structure of all-plus amplitudes. We introduce a convenient diagrammatic formalism that allows us
to deduce relations between these amplitudes, including some special triple-photon decoupling relations.
Section \ref{oneminusSection} is devoted to one-minus amplitudes; here we propose that relations follow from a quite remarkable link to both tree-level and one-loop all-plus amplitudes. In section \ref{moreSection} we will discuss some observations regarding additional relations. Finally, section \ref{conclusion} contains our conclusions and an outlook.

\section{Review of tree-level and one-loop amplitude relations} \label{AmplPropsSection}
The standard color decomposition for $n$-point gauge-theory tree amplitudes of adjoint particles is given by
\begin{align}
\mathcal{A}_n^{\mathrm{tree}} = g^{n-2} \sum_{\sigma\in S_n/\mathbb{Z}_n} \mathrm{Tr}[T^{a_{\sigma(1)}}
\cdots T^{a_{\sigma(n)}}]  A_n^{\mathrm{tree}}(\sigma(1),\ldots,\sigma(n))\,,
\label{colorDec}
\end{align}
where $A_n^{\mathrm{tree}}$ are color-ordered amplitudes, $T^{a_i}$ are $SU(N_c)$ gauge group generators, 
$g$ is the coupling constant, and $S_n/\mathbb{Z}_n$ stands for all non-cyclic permutations. 

The color-ordered amplitudes satisfy several useful identities. First, they are invariant under cyclic 
shifts of the arguments, and similarly up to a sign invariant under reflections, 
\begin{align}
A_n^{\mathrm{tree}}(1,2,\ldots,n) &= A_n^{\mathrm{tree}}(2,3,\ldots,n,1)\,, \nonumber \\  
A_n^{\mathrm{tree}}(1,2,\ldots,n) &=(-1)^n A_n^{\mathrm{tree}}(n,n-1,\ldots,1)\,.
\label{candr}
\end{align}
Second, the amplitudes satisfy special decoupling relations inherited from the gauge group structure
via the notion of color ordering. 
In particular the `photon decoupling' relation takes a simple form,
\begin{align}
0={}& A_n^{\mathrm{tree}}(1,2,3,\ldots,n)+A_n^{\mathrm{tree}}(2,1,3,\ldots,n)+
A_n^{\mathrm{tree}}(2,3,1,\ldots,n)+ \cdots\nn \\ & \hskip7cm \cdots+A_n^{\mathrm{tree}}(2,3,\ldots,1,n)\,.
\label{photonDec}
\end{align}
This expression can be obtained from the decomposition~\eqref{colorDec} where particle 1 is converted to a $U(1)$ field (photon)  
by setting $T^{a_1}\rightarrow 1$. The `photon' must decouple from the remaining adjoint particles, hence the name.
The more general version of~\eqref{photonDec} is given by the Kleiss--Kuijf (KK) relations~\cite{KK,DelDuca}
\begin{align}
A_n^{\mathrm{tree}}(1,\{\alpha\},n,\{\beta\}) ~=~
(-1)^{n_{\beta}} \!\!\!\!\sum_{\sigma\in \mathrm{OP}(\{\alpha\},
\{\beta^T\})}\!\!\!\! A_n^{\mathrm{tree}}(1,\{\sigma\},n)\,,
\label{KKformula}
\end{align}
where the sum is over ``ordered permutations'' OP, that is, all permutations
of $\{\alpha\}\cup \{\beta^T\}$ that maintains the order of the individual elements of each set.
Here $n_{\beta}$ is the number of elements in $\{\beta\}$, and $\{\beta^T\}$ is the $\{\beta\}$ set
with the ordering reversed. From the structure of the right-hand-side of~\eqref{KKformula}, 
it is clear that any tree amplitude can be re-expressed in terms of no more than $(n-2)!$ different 
color-ordered amplitudes. For later use, it is worthwhile to recall that if one represents the 
color-ordered amplitudes in terms of diagrams with only anti-symmetric three-point 
vertices, the KK-relations are automatically satisfied~\cite{BCJ}.
This idea of finding a diagrammatic representation through which relations become 
apparent will be utilized in this paper to derive relations at the one-loop level.

Finally, color-ordered tree amplitudes also satisfy the BCJ relations, worked out by Bern, 
Carrasco and one of the current authors~\cite{BCJ}.
There are several ways of presenting these, but we will be using the following simple $n$-point form
\begin{align}
0 ={}& s_{12}A_n^{\mathrm{tree}}(1,2,3,\ldots,n) + (s_{12}+s_{23})A_n^{\mathrm{tree}}(1,3,2,4,\ldots,n) \nonumber \\
&+ (s_{12}+s_{23}+s_{24})A_n^{\mathrm{tree}}(1,3,4,2,5,\ldots,n)  + \cdots  \nonumber \\
&+ (s_{12}+s_{23}+s_{24}+\cdots + s_{2(n-1)})
A_n^{\mathrm{tree}}(1,3,4,\ldots,n-1,2,n)\,,
\label{BCJ}
\end{align}
along with all the relations obtained by permutations of $\{ 1,2,\ldots,n\}$ in this equation. Here $s_{ij}=(p_i+p_j)^2$. Interestingly, these relations
appear as the field theory limit of monodromy relations in string theory~\cite{BDVmonodromy}
(see also~\cite{Stieberger:2006te,Stieberger,Boels:2010bv} and~\cite{Mafra:2009bz,Tye:2010dd,BjerrumBohr:2010zs} for an alternative 
string theory approach),
but have now also been proven using pure field theory~\cite{Feng:2010my,Chen:2011jx}. The BCJ relations
naturally extend to supersymmetric theories~\cite{Sondergaard:2009za,Jia:2010nz}, a fact that also immediately
follows from the string theory derivation~\cite{BDVmonodromy}.
Combining all such relations reduces the number of independent color-ordered tree amplitudes down to $(n-3)!$

Analogous to tree-level,
the standard color decomposition for $n$-point one-loop amplitudes of adjoint particles is given by
\begin{align}
\mathcal{A}_n^\oneloop ={}& g^n \left[ N_c \sum_{\sigma\in S_n/\mathbb{Z}_n} \mathrm{Tr}[T^{a_{\sigma(1)}}\cdots
T^{a_{\sigma(n)}}]  A_{n;1}^\oneloop(\sigma(1),\ldots,\sigma(n)) \right. \nonumber \\
&\left. \hspace{0.5cm} +\hskip-0.1cm\sum_{c=2}^{\lfloor n/2 \rfloor +1} \hskip-0.2cm\sum_{\sigma\in S_n/S_{n;c}} 
\hskip-0.3cm\mathrm{Tr}[T^{a_{\sigma(1)}}\cdots T^{a_{\sigma({c-1})}}]
\mathrm{Tr}[T^{a_{\sigma(c)}}\cdots T^{a_{\sigma(n)}}] A_{n;c}^\oneloop(\sigma(1),\ldots,\sigma(n)) \right],
\label{loop_decomp}
\end{align}
where $A_{n;1}^\oneloop$ are the planar or \textit{leading}-$N_c$ color-ordered amplitudes, and $A_{n;c>1}^\oneloop$ are the  
\textit{subleading} amplitudes. The $\mathbb{Z}_n$ and $S_{n;c}$ are the subsets of $S_n$ which 
respectively leaves the single and double trace structure invariant, and $\lfloor x \rfloor$ is the largest integer less than 
or equal to $x$.

It is only necessary to consider the amplitudes $A_{n;1}^{(1)}$, since the subleading ones
$A_{n;c>1}^\oneloop$ can all be obtained from
\begin{align}
A_{n;c>1}^\oneloop(1,2,\ldots,c-1,c,c+1,\ldots,n) = (-1)^{c-1}\hspace{-0.5cm}
\sum_{\sigma\in\mathrm{COP}(\{\alpha\}\cup\{\beta\})}\hspace{-0.5cm} A_{n;1}^\oneloop(\sigma(1),\ldots,\sigma(n))\,,
\label{A_c}
\end{align}
where $\{\alpha\} \equiv \{c-1,c-2,\ldots,2,1\}$, $\{\beta\} \equiv \{ c,c+1,\ldots,n-1,n\}$, and $\mathrm{COP}(
\{\alpha\}\cup\{\beta\})$ is
the set of all permutations of $\{\alpha\}\cup\{\beta\}$ that preserve the cyclic ordering of the elements within each set.
These relations follow from enforcing $U(1)$ decoupling in eq.~\eqref{loop_decomp}. Indeed there exist an alternative color decomposition which manifestly incorporates the $U(1)$ decoupling \cite{DelDuca:1999rs},
\begin{align}
\mathcal{A}^\oneloop_n = g^n \sum_{\sigma\in S_{n-1}/\mathcal{R}} &
\mathrm{Tr}[T_{\rm adj}^{a_{\sigma(1)}}\cdots T_{\rm adj}^{a_{\sigma(n)}}]
A_{n;1}^\oneloop(\sigma(1),\ldots,\sigma(n)) \,,
\label{altpure_oneloop}
\end{align}
where $(T_{\rm adj}^a)_{bc} \equiv f^{bac} = \mathrm{Tr}[T^bT^aT^c-T^aT^bT^c]$, $S_{n-1}\equiv S_n/\mathbb{Z}_n$ denotes non-cyclic permutations, which in turn are moded out by reflections $\mathcal{R}$ in the sum.
Since the subleading $A_{n;c>1}^\oneloop$ are not independent quantities, for the rest of this paper we will only consider the $A_{n;1}^\oneloop$ amplitudes. 

For later use we also introduce the corresponding decomposition in a theory with $n_f$ flavors of fundamental-representation quarks, the $n$-gluon one-loop amplitude is extended to \cite{DelDuca:1999rs}
\begin{align}
\mathcal{A'}^{\oneloop}_n = g^n \sum_{\sigma\in S_{n-1}/\mathcal{R}} &
\Big[ \mathrm{Tr}[T_{\rm adj}^{a_{\sigma(1)}}\cdots T_{\rm adj}^{a_{\sigma(n)}}]
A_{n;1}^\oneloop(\sigma(1),\ldots,\sigma(n))  \nonumber \\
& + 2n_f \mathrm{Tr}[T^{a_{\sigma(1)}} \cdots T^{a_{\sigma(n)}}]
A^{[1/2]}_{n;1}(\sigma(1),\ldots,\sigma(n)) \Big]\,,
\label{alt_oneloop}
\end{align}
where $A^{[1/2]}_{n;1}$ are the color-ordered gluon amplitudes with quarks circulating the loop.

The color-ordered amplitudes $A_{n;1}^\oneloop$ and $A^{[1/2]}_{n;1}$ satisfy the exact same cyclic and reflection invariance as the tree-level amplitudes~\eqref{candr}.
However, in general, one might not have guessed the existence of additional amplitude relations at one loop. For example, as already mentioned, the $U(1)$-decoupling identities
have already been used to get eq.~\eqref{A_c}~\cite{ColorDecomp}.
Indeed, as far as we are aware, no explicit sequences of relations that reduce the number of independent amplitudes below 
$(n-1)!/2$ are currently known for generic classes of one-loop amplitudes.

The above setup is valid for any helicity configuration of particles in the adjoint (and fundamental) representation of the gauge group,
for tree and one-loop amplitudes. However, in this paper we will restrict ourself to \textit{finite} one-loop gluonic amplitudes.
That is the $A^{(1)}_{n;1}$ color-ordered amplitudes with configuration of all, or all but one, of the
helicities being equal, \textit{e.g.}  $(+++\cdots+)$ or $(-++\cdots+)$. Before we investigate these amplitudes it is useful to prepare ourselves by looking at the general patterns of the discussed (and expected) amplitude relations.

\subsection{General structure of amplitude relations} \label{AmplStructureSection}

The amplitude properties and relations at tree level have in common that they do not depend on the precise details of the gauge theory, they are rather generic, universal, properties. Details such as gauge group, helicities and particle types are not important in these relations; the key information is the position or the order of the external states in the planar amplitudes. As we seek to generalize such structures of relations to the one-loop level it is convenient to classify the different types of identities we expect to encounter.
 
The simplest type of generic amplitude relations are those with constant numerical coefficients,
\begin{align}
0=\sum_{\sigma} j_{\sigma}\,A_n(\sigma(1),\sigma(2),\ldots,\sigma(n))\,,
\label{KKlikeRel}
\end{align}
where $j_{\sigma}$ are integers, depending only on the permutation $\sigma$. The amplitudes $A_n$ can here be a tree or loop amplitude, and in the latter case it can be leading or subleading color, $A_{n;c}^\oneloop$ or even $A^{[1/2]}_{n;1}$. Examples of such relations are the cyclic and reversal relations \eqref{candr}, photon decoupling relation~\eqref{photonDec}, the KK relations \eqref{KKformula} and \eqref{A_c}. As pointed out in~\cite{BCJ} such relations with constant coefficients can be explained simply from the symmetry properties of underlying diagram expansions. Specifically, tree-level graphs with totally antisymmetric cubic vertices automatically satisfy all tree-level amplitude relations with constant numerical coefficients. In the next section, we will find similar KK-like relations arising at the one-loop level using a diagrammatic representation with particular types of vertices. Alternatively, tree-level relations with constant numerical coefficients can be explained from string theory using monodromies of the world-sheet~\cite{BDVmonodromy}. Remarkably the field theory limit of the real part of these relations can be identified with the KK relations.

The second type of generic amplitude relations are those with coefficients that are products of the kinematical invariants,
\begin{align}
0=\sum_{\sigma}P^{(d)}_{\sigma}(s_{ij})\,A_n(\sigma(1),\sigma(2),\ldots,\sigma(n))\,,
\label{BCJlikeRel}
\end{align}
where $P^{(d)}_{\sigma}(s_{ij})$ are homogeneous polynomials of degree $d$, depending on all the momentum invariants $s_{ij}$. Examples of such relations are the BCJ relations with $d=1$ in~\eqref{BCJ}, or $d\ge1$ in~\cite{BCJ}. At tree-level these relations can be understood from a diagrammatic point of view as arising from nontrivial Jacobi relations between numerators of kinematical diagrams~\cite{BCJ}. Alternatively, using again the monodromy relations from string theory, all such relations at tree-level can be explained as the imaginary part of the monodromy relations~\cite{BDVmonodromy}. For loop-level amplitudes the diagrammatic Jacobi relations have been generalized~\cite{BCJloops}, although corresponding amplitude relations at loop 
level have as of yet not been derived.

\section{KK-like relations for the all-plus-helicity amplitudes} \label{allplusSection} 
We begin the study of relations at the one-loop level by considering the all-plus helicity gluon amplitudes in Yang-Mills theory, written down 
for all multiplicities in~\cite{BernAllPlus},
\begin{align}
A_{n;1}^\oneloop(1^+,2^+,3^+,\ldots,n^+)=-\frac{i}{48\pi^2} \frac{\sum_{i<j<k<l}{\rm Tr}_{-}[ijkl]}{\spa{1}.{2}\spa{2}.{3}\spa{3}.{4}\ldots \spa{n}.{1}}\,,
\label{AllPlusFormula}
\end{align}
where ${\rm Tr}_{-}[ijkl]\equiv \frac{1}{2} {\rm Tr}[(1-\gamma_5){\not}{p}_i {\not}p_j {\not}p_k {\not}p_l]=
\spa{i}.{j}\spb{j}.{k}\spa{k}.{l}\spb{l}.{i}$. The angle and square brackets are standard spinor-helicity notation (see {\it e.g.}~\cite{DixonTASI}).

These amplitudes are some of the simplest nontrivial one-loop objects that one can consider. As such they provide convenient 
testing ground for finding interesting one-loop properties that may generalize to more intricate amplitudes. One remarkable 
feature of the all-plus amplitudes is that they have been conjectured to be related to maximally-helicity-violating (MHV) 
one-loop amplitudes of ${\cal N}=4$ super-Yang-Mills theory through a dimension-shifting formula~\cite{DimShift}. 
Another interesting property is that the full color-dressed all-plus amplitude for $n>4$, in a theory also including quarks, vanish if three or more gluons are replaced by
photons~\cite{BernAllPlus}. We will return to these properties below.

To get started with some explicit examples, let us consider the four-point one-loop all-plus amplitude,
\begin{align}
A_{4;1}^\oneloop(1^+,2^+,3^+,4^+) = -\frac{i}{48\pi^2} \frac{\spb{1}.{4}\spb{2}.{3}}{\spa{1}.{4}\spa{2}.{3}} \,.
\label{4ptallplus}
\end{align}
Using momentum conservation, it is straightforward to see that eq.~\eqref{4ptallplus} is totally crossing symmetric, {\it e.g.}
\begin{align}
 \frac{\spb{1}.{4}\spb{2}.{3}}{\spa{1}.{4}\spa{2}.{3}}= \frac{\spb{1}.{2}\spb{3}.{4}}{\spa{1}.{2}\spa{3}.{4}} = \frac{\spb{1}.{3}\spb{2}.{4}}{\spa{1}.{3}\spa{2}.{4}}\,.\label{4plussym}
\end{align}
From the point of view of the discussion in section~\ref{AmplPropsSection} this property is surprising, as only the cyclic and reversal symmetries are strictly necessary for one-loop amplitudes. To make use of the property (\ref{4plussym}),
we introduce a diagrammatic notation.   
The four-point amplitude will be represented by a single totally symmetric quartic vertex $D^{abcd}$, for instance
\begin{align}
A_{4;1}^\oneloop(1, 2, 3, 4)=D^{a_1a_2a_3a_4}\,, \quad  A_{4;1}^\oneloop(1, 3, 4, 2)=D^{a_1a_3a_4a_2}\,, \quad A_{4;1}^\oneloop(1, 4, 2, 3)=D^{a_1a_4a_2a_3}\,.
\label{4ptdiag}
\end{align}
For notational simplicity, we have suppressed the explicit plus-helicity labels;  in the remaining part of this section we will frequently do the same. The vertex $D^{abcd}$ should be interpreted as a formal object, the explicit indices $a_i$ have no literal meaning (the purpose of these will be explained below). With the help of these  $D^{abcd}$'s the observed symmetry has become manifest.

\FIGURE{
\centerline{\epsfxsize 3.2 truein \epsfbox{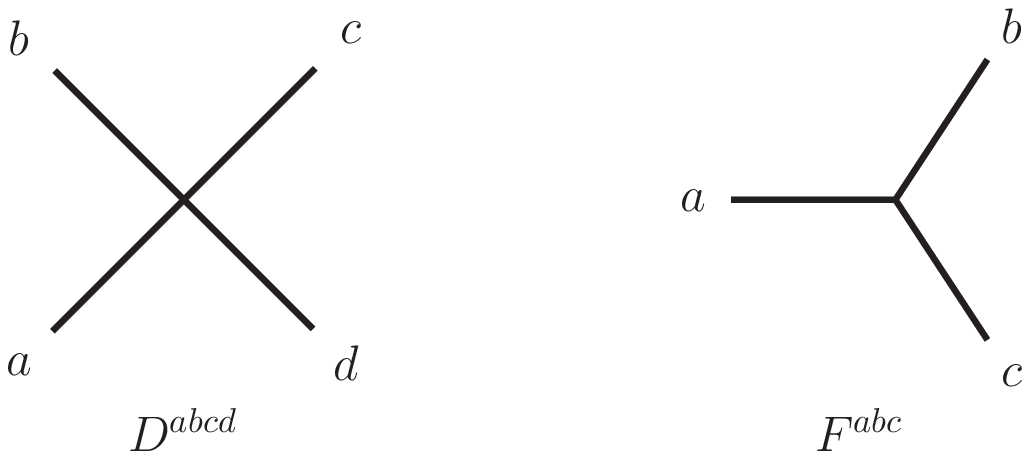}} \caption{Formal vertices used to construct kinematical diagrams for the one-loop all-plus amplitudes.} \label{FormalVerticesFigure}
}

Thus far this has been a completely trivial exercise, but as we go beyond four points we will see that 
this diagrammatic notation becomes surprisingly powerful.  Starting at five points we will introduce an additional formal vertex,
this time a cubic totally antisymmetric vertex $F^{abc}$. We claim that the all-plus one-loop amplitudes 
can be understood as being built  out of diagrams that use these two types of vertices, similar to how only
cubic diagrams can explain tree-level properties of amplitudes~\cite{DelDuca,BCJ}.  Furthermore, we will insist that 
one-loop diagrams have exactly one $D^{abcd}$ vertex, and the remaining ones are of the $F^{abc}$ type.  

\FIGURE{
\centerline{\epsfxsize 4.2 truein \epsfbox{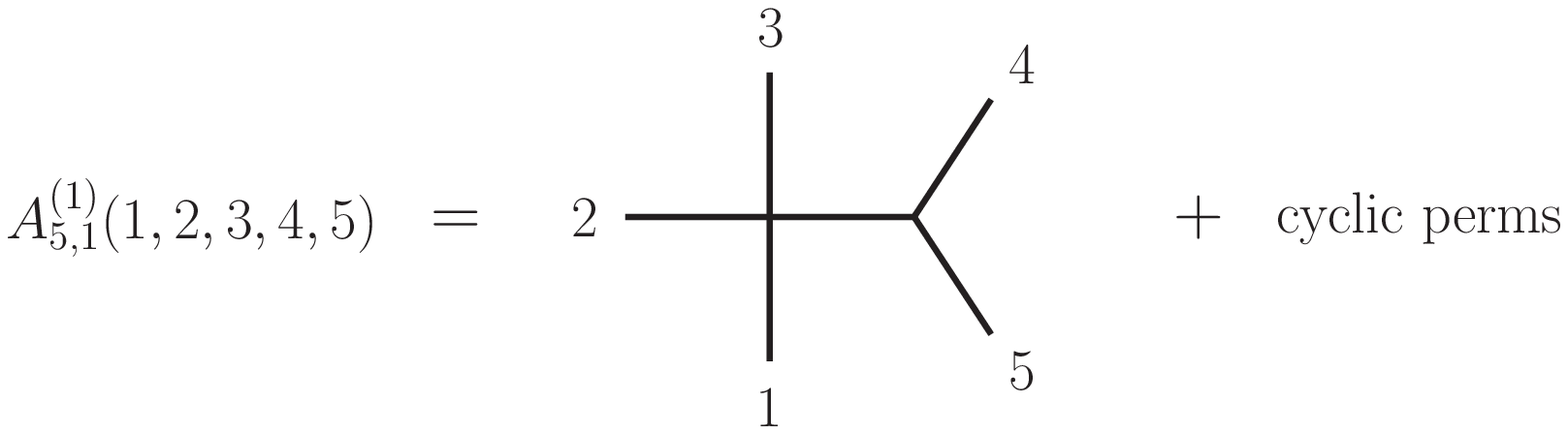}} \caption{Diagram representation of the finite one-loop all-plus five-point amplitude.} \label{FivePointDiagsFigure}
}

For the planar five-point amplitude $A_{5;1}^\oneloop$ this corresponds to the following representation:
\begin{align}
A_{5;1}^\oneloop(1, 2, 3, 4,5)={}& D^{a_1a_2a_3b}F^{b a_4 a_5}+D^{a_2a_3a_4b}F^{b a_5 a_1}
+D^{a_3a_4a_5b}F^{b a_1 a_2} \nonumber \\
& +D^{a_4a_5a_1b}F^{b a_2 a_3}+D^{a_5a_1a_2b}F^{b a_3 a_4}\,.
\label{5ptdiag}
\end{align}
This  diagrammatic expansion is shown in Figure~\ref{FivePointDiagsFigure}.  We stress that the formal vertices  
$D^{abcd}$ and $F^{abc}$ build up the diagrams which in turn are taken to be real objects that incorporate all kinematical 
behavior of the color-stripped amplitudes. In particular, the three-point vertex  $F^{abc}$ is not a gauge group 
structure constant but rather a kinematical analog. The purpose of the indices of these formal kinematical vertices is 
to indicate the contractions, or connectivity, of the vertices, and the position of the different external legs in the diagrams. (In addition to the vertices, one can introduce propagators connecting the vertices, however, for the purposes of this paper it is not needed.) 

There are several ways to motivate this diagrammatic technique. One way is 
to look at the dimension-shifting formula of ref.~\cite{DimShift}. As shown there, 
the all-plus amplitudes in $4-2\epsilon$ dimensions are related to the MHV amplitudes of ${\cal N}=4$
super-Yang-Mills in $8-2\epsilon$ dimensions. Suppressing overall irrelevant factors, for infinitesimal $\epsilon$ we have
\begin{align}
A_{n;1}^\oneloop(1^+,2^+,3^+,\ldots,n^+) \, \sim \, \epsilon \, \langle 12\rangle^{-4}A_{n;1}^{\oneloop,{\cal N}=4}(1^-,2^-,3^+,\ldots,n^+)\Bigl|_{D=8-2\epsilon}
\end{align}
This is illustrated on the cubic diagrams in Figure~\ref{DimShiftingFigure}. The scalar box diagram have an ultraviolet pole $1/ \epsilon$ in $8-2\epsilon$ dimensions that cancels the overall $\epsilon$, resulting in a finite tree-like diagram with a quartic vertex. However, higher-polygon loop diagrams are expected to be finite in this dimension, and after multiplication by $\epsilon$, they vanish as we take $\epsilon \rightarrow 0$. Similarly, triangles and bubbles are not expected to contribute because of the no-triangle property of  the ${\cal N}=4$ theory~\cite{MHVoneloop}. The only surviving diagrams are of the types used in eq.~\eqref{5ptdiag}.

\FIGURE{
\centerline{\epsfxsize 3.5 truein \epsfbox{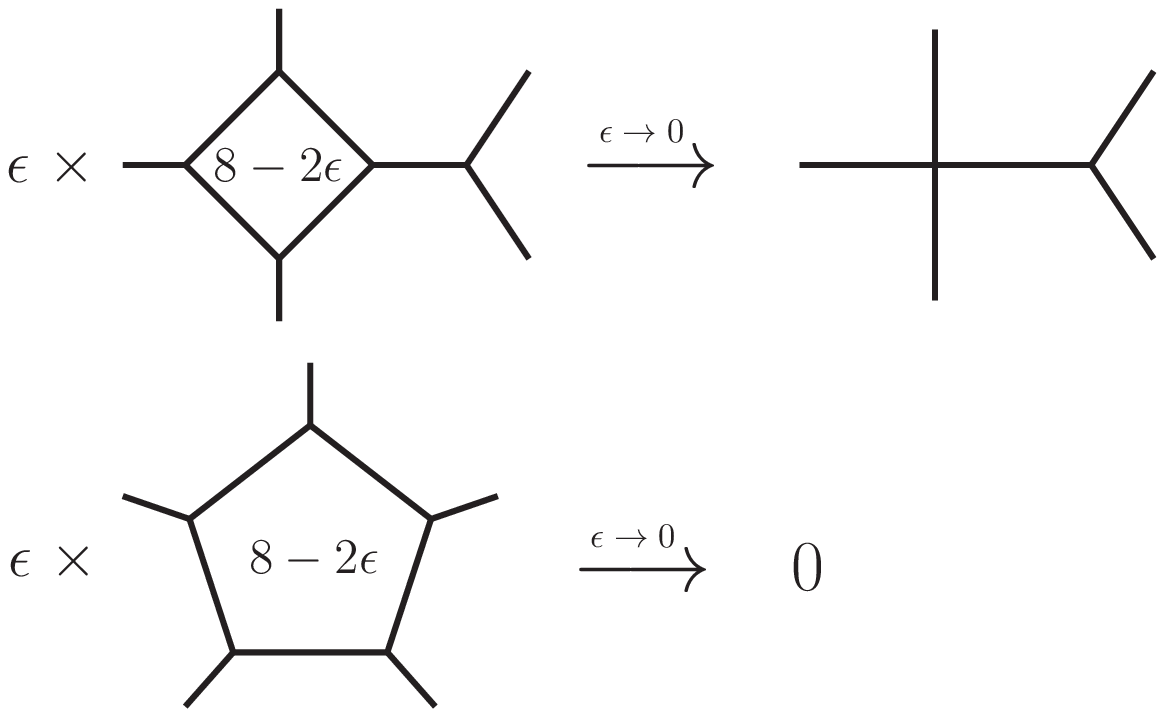}} \caption{The diagrams with quartic vertices can be understood as arising from the dimension shifting relation, in combination with a cubic loop-diagram representation. The box diagrams give finite contributions and higher polygons vanish.} \label{DimShiftingFigure}
}

Now we study the properties of the expression \eqref{5ptdiag}. The cyclicity of this amplitude has become manifest since we sum over all cyclic permutations. A non-trivial consequence of the diagrammatic expansion
is the automatic antisymmetry under reversals or flips. For example,
\begin{align}
A_{5;1}^\oneloop(1, 2, 3, 4,5)={}& -D^{a_1a_2a_3b}F^{b a_5 a_4} -D^{a_2a_3a_4b}F^{b a_1 a_5}
-D^{a_3a_4a_5b}F^{b a_2 a_1} \nonumber \\
& -D^{a_4a_5a_1b}F^{b a_3 a_2}-D^{a_5a_1a_2b}F^{b a_4 a_3} \nonumber \\
={}&-\big( D^{a_3a_2a_1b}F^{b a_5 a_4} +D^{a_4a_3a_2b}F^{b a_1 a_5}
+D^{a_5a_4a_3b}F^{b a_2 a_1} \nonumber \\
& \hspace{0.5cm}+D^{a_1a_5a_4b}F^{b a_3 a_2}+D^{a_2a_1a_5b}F^{b a_4 a_3} \big) \nonumber \\
={}& - A_{5;1}^\oneloop(5,4,3,2,1)\,.
\end{align}
For higher $n$-point amplitudes this antisymmetry under flips (or symmetry, for even $n$)  
immediately generalizes using the diagrammatic representations. 

In order to go beyond five points, we use the precise rules:
For any all-plus one-loop amplitude we represent it as a sum over all distinct planar diagrams
with unit relative weight, such that each diagram contains
exactly \textit{one} totally symmetric quartic vertex $D^{abcd}$ and all remaining legs are
attached by means of anti-symmetric cubic vertices $F^{abc}$ in all possible planar distinct ways.

For six points, this gives the diagrammatic expansion shown in Figure~\ref{SixPointDiagsFigure}. 
In detail,
\begin{align}
A_{6;1}^\oneloop(1, 2, 3, 4,5,6)={}& D^{a_1a_2a_3b}F^{b a_4 c}F^{c a_5 a_6}
+D^{a_1a_2a_3b}F^{b c a_6}F^{c a_5 a_4}+D^{a_1a_2bc}F^{b a_3 a_4}F^{c a_5 a_6} \nonumber \\
&+\frac{1}{2}F^{a_1 a_2b}D^{ba_3ca_6}F^{c a_4 a_5} ~+~{\rm cyclic}\,,
\label{6ptdiag}
\end{align}
where ``cyclic'' instructs us to sum over all cyclic permutations of $\{1,2,3,4,5,6\}$. The factor of 1/2 is inserted in order to compensate for the over-counting of the fourth diagram in this cyclic sum. In total there are 21 contributing diagrams in $A_{6;1}^\oneloop(1, 2, 3, 4,5,6)$, see Table~\ref{NPointTable}.

\FIGURE{
\centerline{\epsfxsize 5.4 truein \epsfbox{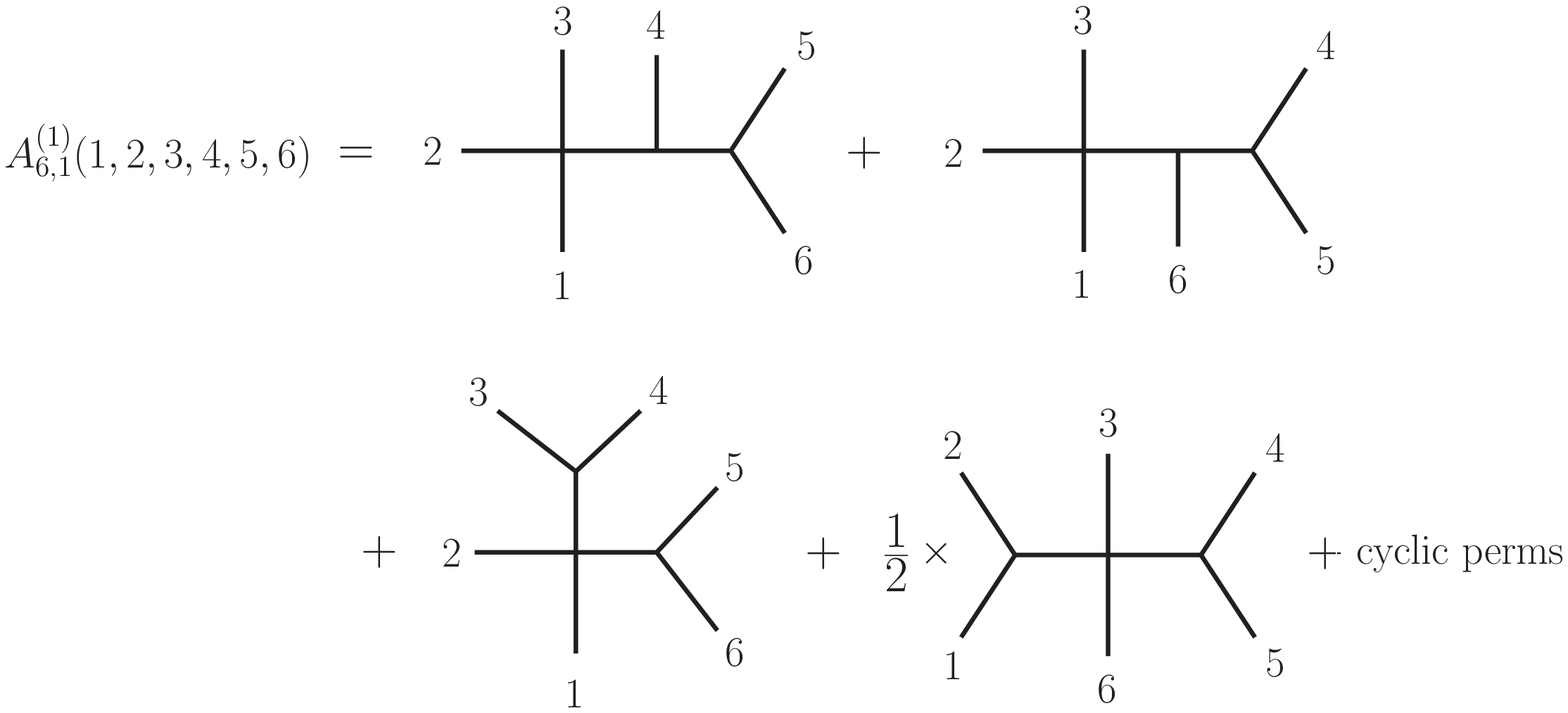}} \caption{Diagram representation of the finite one-loop all-plus six-point amplitude.} \label{SixPointDiagsFigure}
}

As illustrated by these simple examples, one can immediately build up a diagrammatic representation
for any all-plus-helicity one-loop amplitude. In~Table~\ref{NPointTable} we have worked out the $n$-point case in all generality. As shown the total number of diagrams grows faster than the number of color-ordered amplitudes. Nonetheless, the diagrammatic structure imposes severe constraints on these amplitudes. Using {\it only} the properties that the quartic vertex is totally symmetric and that the cubic vertex is totally antisymmetric, one obtains relations between the various amplitudes. For example, at five and six points the relations reduce the number of independent amplitudes as $12\rightarrow 6$ and $60\rightarrow35$, respectively. Indeed, the amplitudes, as given by \eqref{AllPlusFormula}, satisfy all relations we obtain from the diagram symmetry properties.

For general $n$, after taking into account all KK-like relations~\eqref{KKlikeRel}, the number of independent amplitude curiously coincides with a particular sequence of the unsigned Stirling numbers of the first kind, denoted by $c(n-1,3)$. These numbers have a particular nice combinatorial interpretation. The $c(n-1,3)$ sequence correspond to the number of ways $n-1$ people can be seated at three round tables, with at least one person per table. Using the diagrammatic representation this count for the  KK-independent amplitudes has been explicitly confirmed up to eight points. In section~\ref{moreSection} we will see that the number of independent amplitudes can be further reduced using BCJ-like amplitude relations~\eqref{BCJlikeRel}.

\def\hs{\hskip .2 cm \null }
\begin{table*}
\caption{The counting of diagrams, amplitudes and KK-like relations at different $n$, for the all-plus-helicity amplitudes. The diagrams have exactly one quartic vertex, and the remaining ones are cubic. `KK-like~relations' are those with constant coefficients, the BCJ-like relations are not included in these counts.  `KK-indep.~ampls.'  are the number of amplitudes independent under the KK-like relations. The $c(n,k)$ are the unsigned Stirling numbers of the first kind. All formulas have been explicitly checked up to at least $n=15$, except for the last two rows, which have been checked to eight points.}
\label{NPointTable} 	
\vskip .4 cm
\begin{tabular}{|l|*{6}{c}|c|}
\hline
external legs  	& $\phantom{i}$ 4 $\phantom{i}$ &  $\phantom{x}$ 5$\phantom{x}$&  $\phantom{x}$ 6$\phantom{x}$& $\phantom{xx}$  7  $\phantom{xx}$ &  $\phantom{xx}$ 8 $\phantom{xx}$ &  $n$ \\
\hline

all diagrams                     & 1 & 10 & 105 & 1260 & 17325 &  $\frac{(2 n - 5)!}{6 (n - 4)! 2^{n - 4}}$ \\
amplitudes $A_{n;1}$    & 3 & 12 & 60 & 360 &  2520  & $\frac{1}{2}(n-1)!$ \\
diagrams in $A_{n;1}$   & 1 & 5 & 21 & 84 & 330 &  $\frac{(2n - 5)!}{(n - 4)! (n - 1)!}$ \\
KK-like relations              & 2 & 6 & 25 & 135 &  896 & $\frac{1}{2}(n-1)!-c(n-1,3)$  \\
KK-indep. ampls.            & 1 & 6 & 35 &  225 &  1624  &  $c(n-1,3)$\\
\hline
\end{tabular}
\vskip .5 cm
\end{table*}

\subsection{Explicit all-plus amplitude relations}
In this section we will write down some of the concrete relations for the all-plus-helicity amplitudes. The relations are in general quite intricate (compared to the simplicity of~\eqref{AllPlusFormula}) so we will not elaborate on every explicit case. Instead, we will focus on some of the interesting patterns that we will return to in the sections below.

Starting at four points, using eq. (3.4), the amplitude is crossing symmetric by construction. This gives the trivial relations
\begin{align}
A_{4;1}^\oneloop(1,2,3,4) = A_{4;1}^\oneloop(1,3,4,2) = A_{4;1}^\oneloop(1,4,2,3)\,.
\end{align}
Next at five points, it follows from eq.~\eqref{5ptdiag} that the following simple relation must be satisfied:
\begin{align}
0 ={}& A_{5;1}^\oneloop( 1, 4, 3, 5,2) + A_{5;1}^\oneloop(1, 5, 3, 4,2) +
 A_{5;1}^\oneloop(1,2, 3, 4, 5) \nonumber \\
&+ A_{5;1}^\oneloop(1,2, 3, 5, 4) +
 A_{5;1}^\oneloop(1, 5, 3,2,4) + A_{5;1}^\oneloop(1, 4, 3,2,5)\,.
\label{5ptKKlike}
\end{align}
For example, we can study how the diagram $D^{a_1a_2a_3b}F^{b a_4 a_5}$ enters into this relation. Of the six above amplitudes only $A_{5;1}^\oneloop(1,2, 3, 4, 5)$ and $A_{5;1}^\oneloop(1,2, 3, 5, 4)$ contains this diagram. Since $F^{b a_4 a_5}$ is 
antisymmetic in $4\leftrightarrow5$ the diagram indeed cancels out in the sum of these two amplitudes. Similar analysis 
shows that any one of the possible ten diagrams one can construct at the five-point level cancels out in~\eqref{5ptKKlike}. 
Using the explicit expression for the five-point amplitudes \eqref{AllPlusFormula} we indeed find that~\eqref{5ptKKlike} 
is a true relation. This shows that our diagrammatic representation correctly captures these properties of the amplitudes.

To further elaborate on the five-point case, we can write \eqref{5ptKKlike} more compactly as
\begin{align}
0 ={}&  2A_{5;1}^\oneloop(1,2,3,4,5) + (-1)^{2}\hspace{-0.8cm}\sum_{\sigma\in \mathrm{OP}(\{4\} \cup \{2,1\})}
\hspace{-0.8cm} A_{5;1}^\oneloop(3,\{\sigma\},5) \nonumber \\
& + \mathcal{P}(4,5)\,,
\label{5ptgenKK}
\end{align}
where $\mathcal{P}(4,5)$ is just the same expression as in the first line, but with leg 4 and 5 interchanged. Note
that the second term of the first line takes the same form as a KK-relation for an amplitude
of ordering $(1,2,3,4,5)$. This formula can be generalized to arbitrary $n$ by means of
\begin{align}
0 ={}&  2A_{n;1}^\oneloop(1,2,\ldots,n) + (-1)^{n-3}\hspace{-0.5cm}\sum_{\sigma\in \mathrm{OP}(\{4\} \cup 
\{\beta\})}\hspace{-0.5cm}
 A_{n;1}^\oneloop(3,\{\sigma\},5) \nonumber \\
& +\mathcal{P}(4,5,\ldots,n)\,,
\label{nptKKgen}
\end{align}
with $\{\beta\} = \{ 2,1,n,n-1,\ldots,6\}$. We have explicitly verified this relation up to $n=20$
using the analytical expression for the 
amplitudes~\eqref{AllPlusFormula}. The diagrammatic approach generates several other structures not 
included in~\eqref{nptKKgen}. As a particular important example, consider the following relation at five points:
\begin{align}
0 ={}& 3 A_{5;1}^\oneloop(1, 2, 3, 4, 5) - A_{5;1}^\oneloop(1, 2, 3, 5, 4) -
 A_{5;1}^\oneloop(1, 2, 4, 3, 5) - A_{5;1}^\oneloop(1, 2, 4, 5, 3) \nonumber \\
&{}-  A_{5;1}^\oneloop(1, 2, 5, 3, 4) - A_{5;1}^\oneloop(1, 3, 2, 4, 5) -
 A_{5;1}^\oneloop(1, 3, 4, 2, 5) - A_{5;1}^\oneloop(1, 3, 4, 5, 2) \nonumber \\
&{}- A_{5;1}^\oneloop(1, 4, 2, 3, 5) - A_{5;1}^\oneloop(1, 4, 2, 5, 3) -
 A_{5;1}^\oneloop(1, 4, 5, 2, 3) - A_{5;1}^\oneloop(1, 5, 2, 3, 4)\,.
\label{5ptQ3}
\end{align}
This can be written in a more compact way as
\begin{align}
0 = 6A_{5;1}^\oneloop(1,2,3,4,5) -  \sum_{k=2}^{4}\left[ \sum_{\sigma_k\in\mathrm{OP}(\{\alpha_k\}\cup \{\beta_k\})}
\hspace{-0.5cm} A_{5;1}^\oneloop(1,\{\sigma_k\}) \right]\,,
\label{5ptgen}
\end{align}
where $\{ \alpha_k\} \equiv \{2,3,\ldots,k\}$, $\{ \beta_k\} \equiv \{k+1,\ldots,5\}$ and OP is the set of 
``ordered permutations''  obtained from the two sets. For instance, for $k=3$ we have $\{ \alpha_3\} \equiv \{2,3\}$ and 
$\{ \beta_3\} \equiv \{4,5\}$, so
the $\sigma_3$ sum runs over the permutations
\begin{align}
\{2,3,4,5\}, \quad \{2,4,3,5\}, \quad \{2,4,5,3\}, \quad \{4,2,3,5\}, \quad \{4,2,5,3\}, \quad \{4,5,2,3\}.
\end{align}
This generalizes immediately to $n$-point amplitudes in the following way:
\begin{align}
0 = 6A_{n;1}^\oneloop(1,2,\ldots,n) - \sum_{k=2}^{n-1}\left[ \sum_{\sigma_k\in\mathrm{OP}(\{\alpha_k\}\cup \{\beta_k\})}
\hspace{-0.5cm} A_{n;1}^\oneloop(1,\{\sigma_k\}) \right],
\label{allplus3}
\end{align}
where $\{ \alpha_k\} \equiv \{2,3,\ldots,k\}$ and $\{ \beta_k\} \equiv \{k+1,\ldots,n\}$. 
We have explicitly checked that these relations are satisfied up to 15 points
using~\eqref{AllPlusFormula}.

To make \eqref{allplus3} more concrete we may count the number of terms in these identities.
For $n$-point functions there are $2^{n-1}-n+1$ terms, except for the one case of
$n=8$, when there are only $2^{n-1}-n$ terms. The coefficient of the first amplitude $A_{n;1}^\oneloop(1,2,\ldots,n)$ in the identity is $6-(n-2)=(8-n)$, and thus at precisely $n=8$ this term is not included in the
sum. All other coefficients in the sum are unity.

\subsection{Triple-photon decoupling relations}
The all-plus-helicity amplitudes satisfy a particularly interesting type of relations which we refer to as `triple-photon decoupling'~\cite{BernAllPlus}. These relations are included in the general count of KK-like relations in Table~\ref{NPointTable}. Let us examine them in some detail. First we note that photons or $U(1)$ particles always decouple from the full amplitude if no fundamental particles are present, as discussed in section~\ref{AmplPropsSection}. However, the triple-photon decoupling relations we treat here are in fact non-trivial relations when applied to the color-ordered amplitudes,  $A_{n;1}^\oneloop$ or $A^{[1/2]}_{n;1}$.

We consider the all-plus amplitudes, with both adjoints gluons and fundamental quarks in the loop, using $\mathcal{A'}^{\oneloop}_{n}$ in eq.~\eqref{alt_oneloop}. If we replace one or more of the external gluons with photons, then we have $T^{a_i} \rightarrow 1$ and $T_{\rm adj}^{a_i}  \rightarrow 0$. The first line of eq.~\eqref{alt_oneloop} thus vanishes. The second line of eq.~\eqref{alt_oneloop}, however, will just start to collect into sums of $A^{[1/2]}_{n;1}$ with trace factors involving fewer $T^{a_j}$ matrices.

An interesting result found in ref.~\cite{BernAllPlus} is that for $n>4$ and for  three or more external photons the full one-loop amplitude vanishes,  $\mathcal{A'}^{\oneloop}_{n>4} = 0$. Collecting the $A^{[1/2]}_{n;1}$ amplitudes after replacing three or more $T^{a_i}$'s with unity one immediately finds the relations
\begin{align}
0 ~~= \sum_{\sigma\in\mathcal{P}(\mathrm{O}(\{\alpha\})\cup \{\beta\})} \hspace{-0.4cm} A^{[1/2]}_{n;1}(1,\{\sigma\})\,,
\label{unirel_quark}
\end{align}
where we have used cyclicity to fix the position of leg 1, and $\{\alpha\}$ and $\{\beta\}$ is any partition of the legs $\{2,3,\ldots,n\}$ ($n>4$), with $\{\beta\}$ containing at least three elements (the photons). The sum is over all permutations of
leg $\{2,3,\ldots,n\}$ with the order of the elements in the $\{\alpha\}$ set kept fixed.

Moreover, due to the supersymmetric Ward identity $A^{\mathrm{SUSY}}(1^{\pm},2^+,\ldots,n^+) = 0$,
color-ordered amplitudes with quarks circulating the loop are equal, up to a sign, to the color-ordered amplitudes having
only gluons in the loop \cite{Grisaru}. Instead of eq.~\eqref{unirel_quark} we can therefore write
\begin{align}
0 ~~= \sum_{\sigma\in\mathcal{P}(\mathrm{O}(\{\alpha\})\cup \{\beta\})} \hspace{-0.4cm} A_{n;1}^\oneloop(1,\{\sigma\})\,,
\label{tripid}
\end{align}
where the color-ordered amplitudes again only involve gluons in the loop. By a combination of supersymmetric Ward identities and triple-photon decoupling, we thus obtain identities that hold for the all-plus gluon amplitudes, even with no fermions present at all.

We can also understand the triple-photon decoupling of the all-plus amplitudes in terms of the diagrammatic representation introduced in section
\ref{allplusSection}. First we note that converting a single gluon to a $U(1)$ state (photon), \textit{i.e.} taking a $T^{a_i}\rightarrow 1$, implies that we start to collect color-ordered amplitudes into sums where the photon leg appears in all different positions, much like what we see in eq.~\eqref{photonDec}. Thinking of this sum in terms of the color-ordered diagrams the photon leg also here appears in all possible locations. Let us first
assume that the photon leg is attached to a cubic vertex. Two such locations are shown in Figure~\ref{PhotonDecouplingFigure}. In these diagrams the photon leg attaches to the same internal line through a cubic vertex in two different orderings, and the remaining parts of the diagrams are identical (hidden inside the blobs). Using the kinematical $F^{abc}$'s this picture corresponds to the following (suppressing all vertex factors inside the blobs):
\begin{align}
F^{a \gamma b}+F^{a b \gamma}=0\,,
\end{align}
which is nothing but the antisymmetry property of the cubic vertex. Since the diagrams in Figure~\ref{PhotonDecouplingFigure} are  generic tree-like diagrams, and the photon leg will appear in all different places, we conclude that diagrams with the photon leg attached to a cubic vertex will always appear in canceling pairs.

\FIGURE{
\centerline{\epsfxsize 5 truein \epsfbox{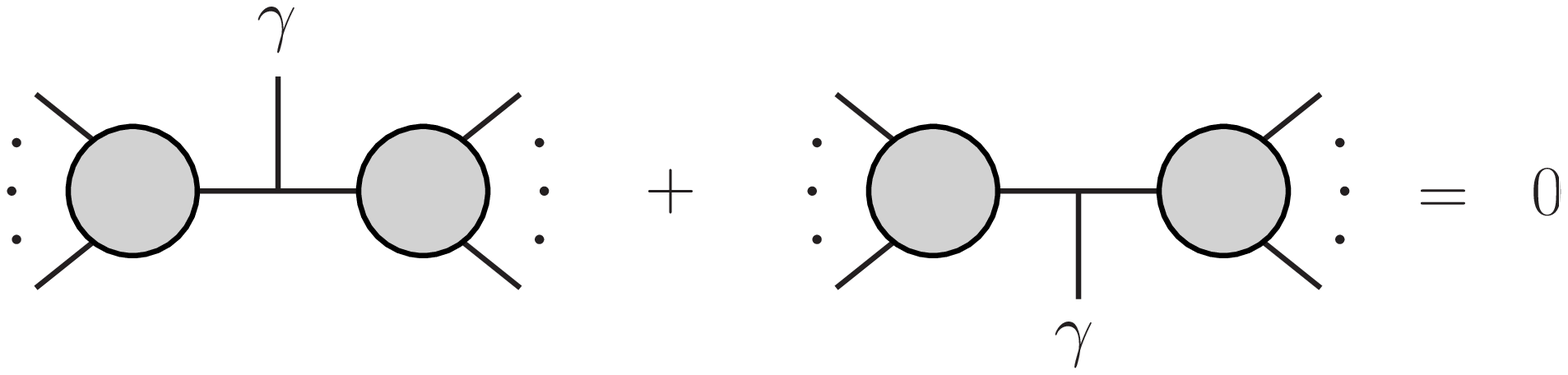}} \caption{ A photon, being a colorless particle, appears in all possible positions in the planar color-ordered diagrams. Because the cubic kinematical vertex is antisymmetric the diagrams will cancel in pairs when photons attach to such vertices.} \label{PhotonDecouplingFigure}
}

For the diagrams where the photon leg is attached to a symmetric quartic vertex the above argument does not apply. Even with two photon legs we will still have nonvanishing diagrams where both photon legs are attached to a quartic vertex. However, if we take three photon legs we observe a new kind of vanishing, illustrated in Figure \ref{TriPhotonDecouplingFigure}. Either at least one photon leg is attached to a cubic vertex, in which case we already know it will cancel with another diagram, or all three photon leg is attached to the quartic vertex. However, since the sum now contains diagrams with the three photon legs in all different positions they will again start to appear in canceling pairs,
\begin{align}
(F^{a c b} +F^{a b c})D^{c \gamma_1 \gamma_2 \gamma_3}=0\,.
\end{align}
Note that the four-point case is an exception because the $F^{abc}$ vertices, needed for the cancelations, are not present in these amplitudes.	

\FIGURE{
\centerline{\epsfxsize 4.7 truein \epsfbox{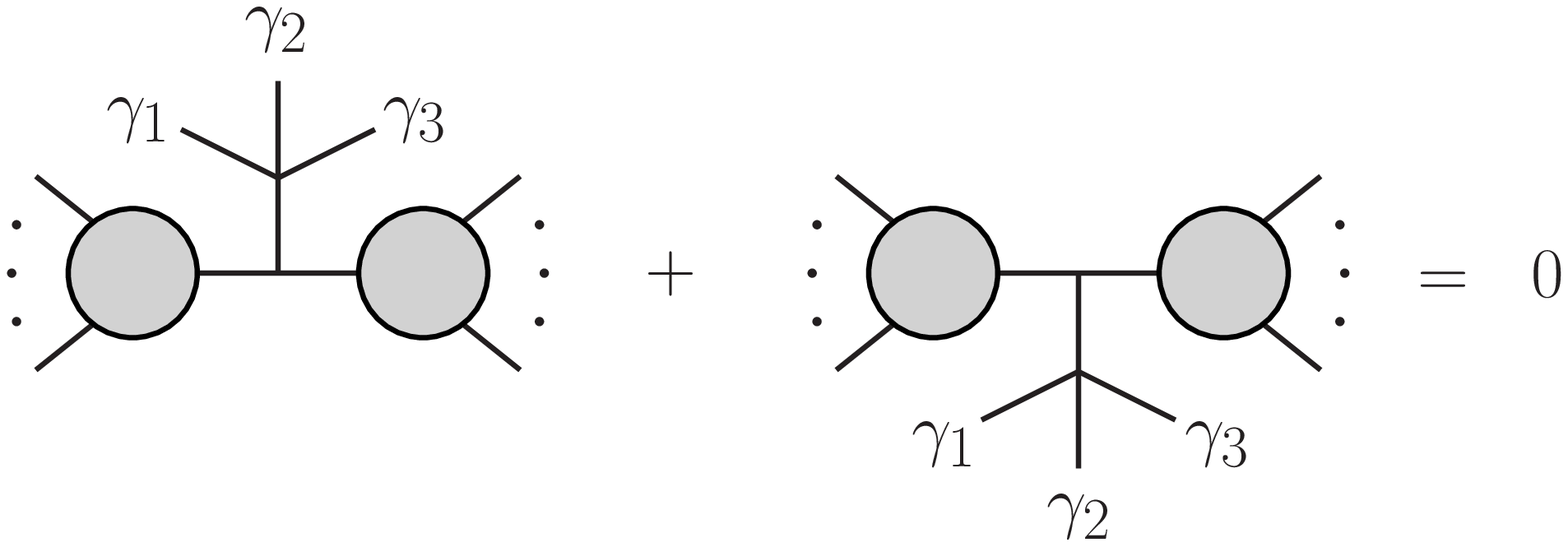}} \caption{Photons can couple to the quartic symmetric vertex present in the diagrams. For one or two photons this give non-vanishing contributions to the color-ordered amplitudes, but for three or more they
again cancel in pairs, as illustrated.
 \label{TriPhotonDecouplingFigure}}
}

To illustrate triple-photon decoupling identities of the all-plus amplitudes, let us give some examples using \eqref{tripid}.
At five points the choices $\{\alpha\} = \{2\}$, $\{\beta\}=\{3,4,5\}$, or $\{\alpha\} = \emptyset$, $\{\beta\}=\{2,3,4,5\}$
result in trivial relations since all $(n-1)!$ permutations are summed over, and the amplitudes will just appear in canceling pairs
$A_n(1,\sigma) + A_n(1,\sigma^T)$, by the trivial reversal antisymmetry \eqref{candr}. 

For $n\ge6$ the relations will be nontrivial. For instance, at six points with $\{\alpha\} = \{2,3\}$ and $\{\beta\}=\{4,5,6\}$ we get
\begin{align}
0={}& A_{6;1}^\oneloop(1, 2, 3, 4, 5, 6) + A_{6;1}^\oneloop(1, 2, 3, 4, 6, 5) + A_{6;1}^\oneloop(1, 2, 3, 5, 4, 6) + A_{6;1}^\oneloop(1, 2, 3, 5, 6, 4) \nonumber \\
{}& + A_{6;1}^\oneloop(1, 2, 3, 6, 4, 5) +A_{6;1}^\oneloop(1, 2, 3, 6, 5, 4) + A_{6;1}^\oneloop(1, 2, 4, 3, 5, 6) + A_{6;1}^\oneloop(1, 2, 4, 3, 6, 5) \nonumber \\
{}& + A_{6;1}^\oneloop(1, 2, 4, 5, 3, 6) + A_{6;1}^\oneloop(1, 2, 4, 5, 6, 3) +A_{6;1}^\oneloop(1, 2, 4, 6, 3, 5) + A_{6;1}^\oneloop(1, 2, 4, 6, 5, 3) \nonumber \\
{}& + A_{6;1}^\oneloop(1, 2, 5, 3, 4, 6) + A_{6;1}^\oneloop(1, 2, 5, 3, 6, 4) + A_{6;1}^\oneloop(1, 2, 5, 4, 3, 6)+A_{6;1}^\oneloop(1, 2, 5, 4, 6, 3) \nonumber \\
{}&+ A_{6;1}^\oneloop(1, 2, 5, 6, 3, 4) + A_{6;1}^\oneloop(1, 2, 5, 6, 4, 3) + A_{6;1}^\oneloop(1, 2, 6, 3, 4, 5) + A_{6;1}^\oneloop(1, 2, 6, 3, 5, 4)\nonumber \\
{}& +A_{6;1}^\oneloop(1, 2, 6, 4, 3, 5) + A_{6;1}^\oneloop(1, 2, 6, 4, 5, 3) + A_{6;1}^\oneloop(1, 2, 6, 5, 3, 4) + A_{6;1}^\oneloop(1, 2, 6, 5, 4, 3) \nonumber \\
{}&+ A_{6;1}^\oneloop(1, 4, 2, 3, 5, 6) +A_{6;1}^\oneloop(1, 4, 2, 3, 6, 5) + A_{6;1}^\oneloop(1, 4, 2, 5, 3, 6) + A_{6;1}^\oneloop(1, 4, 2, 5, 6, 3) \nonumber \\
{}&+ A_{6;1}^\oneloop(1, 4, 2, 6, 3, 5) + A_{6;1}^\oneloop(1, 4, 2, 6, 5, 3) +A_{6;1}^\oneloop(1, 4, 5, 2, 3, 6) + A_{6;1}^\oneloop(1, 4, 5, 2, 6, 3) \nonumber \\
{}&+ A_{6;1}^\oneloop(1, 4, 5, 6, 2, 3) + A_{6;1}^\oneloop(1, 4, 6, 2, 3, 5) + A_{6;1}^\oneloop(1, 4, 6, 2, 5, 3) +A_{6;1}^\oneloop(1, 4, 6, 5, 2, 3) \nonumber \\
{}&+ A_{6;1}^\oneloop(1, 5, 2, 3, 4, 6) + A_{6;1}^\oneloop(1, 5, 2, 3, 6, 4) + A_{6;1}^\oneloop(1, 5, 2, 4, 3, 6) + A_{6;1}^\oneloop(1, 5, 2, 4, 6, 3)\nonumber \\
{}&
 +A_{6;1}^\oneloop(1, 5, 2, 6, 3, 4) + A_{6;1}^\oneloop(1, 5, 2, 6, 4, 3) + A_{6;1}^\oneloop(1, 5, 4, 2, 3, 6) + A_{6;1}^\oneloop(1, 5, 4, 2, 6, 3) \nonumber \\
{}&+ A_{6;1}^\oneloop(1, 5, 4, 6, 2, 3)
 +A_{6;1}^\oneloop(1, 5, 6, 2, 3, 4) + A_{6;1}^\oneloop(1, 5, 6, 2, 4, 3) + A_{6;1}^\oneloop(1, 5, 6, 4, 2, 3) \nonumber \\
{}&+ A_{6;1}^\oneloop(1, 6, 2, 3, 4, 5) + A_{6;1}^\oneloop(1, 6, 2, 3, 5, 4)
 +A_{6;1}^\oneloop(1, 6, 2, 4, 3, 5) + A_{6;1}^\oneloop(1, 6, 2, 4, 5, 3) \nonumber \\
{}&+ A_{6;1}^\oneloop(1, 6, 2, 5, 3, 4) + A_{6;1}^\oneloop(1, 6, 2, 5, 4, 3) + A_{6;1}^\oneloop(1, 6, 4, 2, 3, 5)
 +A_{6;1}^\oneloop(1, 6, 4, 2, 5, 3) \nonumber \\
{}&+ A_{6;1}^\oneloop(1, 6, 4, 5, 2, 3) + A_{6;1}^\oneloop(1, 6, 5, 2, 3, 4) + A_{6;1}^\oneloop(1, 6, 5, 2, 4, 3) + A_{6;1}^\oneloop(1, 6, 5, 4, 2, 3)\, .
\label{uni6pt}
\end{align}
This relation contains all 60 amplitudes that are
not related by cyclicity or reflection. The choices, $\{\alpha\} = \{2\}, \{\beta\} = \{3, 4, 5, 6\}$, and
$\{\alpha\} = \emptyset$, $\{\beta\}=\{ 2,3,4,5,6\}$ also result in
valid relations. It is, however, easy to convince oneself that these are not independent of the
relations where $\{\beta\}$ contains exactly 3 elements.
Indeed they will be just sums of relations of the form given in eq.~\eqref{uni6pt}.

At seven points, the case of $\{\alpha\} = \{2,3,4\}$ and $\{\beta\} = \{5,6,7\}$ leads to a relation containing
120 amplitudes. Again, the remaining relations obtained
from $\{\alpha\}=\{2,3\}$, $\{\beta\}=\{4,5,6,7\}$ and $\{\alpha\}=\{2\}$,
$\{\beta\}=\{3,4,5,6,7\}$ and $\{\alpha\}=\emptyset$, $\{\beta\}=\{2,3,4,5,6,7\}$ are not independent of the
identities where $\{\beta\}$ contains exactly 3 elements, but just sums of relations with this form.

\section{One-minus-helicity amplitude relations\label{oneminusSection}}
We now go on to consider the second class of finite one-loop gluon amplitudes, 
those with one negative helicity leg and the remaining ones being positive $(-++\cdots +)$. 
We will show that there also exist surprising monodromy-like relations for these, both in the class of eq.~\eqref{KKlikeRel}
and eq.~\eqref{BCJlikeRel}.

Although the one-minus amplitudes are finite they have a much more complicated appearance than the 
all-plus amplitudes. The first first few are~\cite{StringBasedRules,BDK1,BDK2}
\begin{align}
A_{4;1}^\oneloop(1^-,2^+,3^+,4^+)=\frac{i}{3}\frac{\spa{2}.{4}\spb{2}.{4}^3}{\spb{1}.{2}\spa{2}.{3}\spa{3}.{4}\spb{4}.{1}}\,,
\label{oneMinusAmp}
\end{align}
and 
\begin{align}
A_{5;1}^\oneloop(1^-,2^+,3^+,4^+,5^+)=\frac{i}{3}\frac{1}{\spa{3}.{4}^2}\left[-\frac{\spb{2}.{5}^3}{\spb{1}.{2}\spb{5}.{1}}+\frac{\spa{1}.{4}^3\spb{4}.{5}\spa{3}.{5}}{\spa{1}.{2}\spa{2}.{3}\spa{4}.{5}^2}-\frac{\spa{1}.{3}^3\spb{3}.{2}\spa{4}.{2}}{\spa{1}.{5}\spa{5}.{4}\spa{3}.{2}^2}\right]\,.
\end{align}
The higher-point amplitudes at any multiplicity were first calculated by Mahlon~\cite{Mahlon} using off-shell recursive techniques. More compact expressions were later obtained by on-shell recursion~\cite{lastfinite}. Although similar to the 
two amplitudes above, the higher-point amplitudes are structurally quite complicated. 
Nonetheless, we will see that interesting relations exist and can be uncovered even in this case.

The treatment of the  one-minus amplitudes will be vastly facilitated by the following observation or principle:
\\
\\
\textit{Amplitude relations which are valid for both tree-level amplitudes and all-plus one-loop amplitudes, are automatically also satisfied for the one-minus one-loop amplitudes.}
\\
\\
That is, using \textit{only} properties of the simple all-plus one-loop amplitudes and the well-understood tree-level amplitudes, we claim to
obtain relations for the more complicated one-minus one-loop amplitudes.

Although we will simply state this as our principle, 
one can again motivate it using diagrams. As observed in \cite{BernAllPlusVertex}, 
at one loop there appears to exist an effective on-shell three-point all-plus-helicity amplitude,
\begin{align}
A_{4;1}^\oneloop(1^+,2^+,3^+) \sim \frac{1}{K^2}\spb{1}.{2}\spb{2}.{3}\spb{3}.{1} \, ,
\label{AllPlusVertex}
\end{align}
where $K^2$ is a scale to get the dimensions right. This amplitude is known to enter the on-shell recursion relations for the one-minus amplitudes and other finite pieces \cite{lastfinite,BernAllPlusVertex}, therefore it is a quite interesting object to consider. In our diagrammatic language this amplitude cannot be identified with the quartic vertex $D^{abcd}$ used previously, nor can it arise from the cubic $F^{abc}$ vertex (which should be identified with tree amplitude contributions). 

We should instead introduce a new cubic effective vertex $\tilde F^{abc}$, which inherits the total antisymmetry from \eqref{AllPlusVertex}. Now, the diagrams we use to represent the one-minus amplitude are the following: sum over all diagrams that have exactly one $D^{abcd}$ or exactly one $\tilde F^{abc}$ vertex, for the remaining vertices use $F^{abc}$.  The two types of diagrams are formally independent, so we can consider them separately. The diagrams with $D^{abcd}$ will by themselves satisfy the same properties and relations as they did in section~\ref{allplusSection}. Similarly, the cubic diagrams containing one $\tilde F^{abc}$ and the rest of the $ F^{abc}$'s will formally behave like cubic tree-level diagrams~\cite{BCJ}. Since for all purposes considered here the two types of cubic vertices satisfy the same properties, namely total antisymmetry. These diagrams should satisfy the same KK-relations as the tree amplitudes. The one-loop one-minus-heliciy amplitudes, being a sum of cubic and quartic diagrams, should then only satisfy those relations that hold simultaneously for both tree amplitudes and all-plus-helicity amplitudes. This is equivalent to the above principle.

\def\hs{\hskip .2 cm \null }
\begin{table*}
\caption{The counting of amplitudes and KK-like relations at different $n$, for the one-minus-helicity amplitudes. The `KK-like~relations' are those with constant coefficients, the BCJ-like relations are not included in these counts.  `KK-indep.~ampls.'  are the number of amplitudes independent under the KK-like relations. The $c(n,k)$ are the unsigned Stirling numbers of the first kind. All entries have been explicitly checked up to at least eight points.}
\label{NPointTable2} 	
\vskip .4 cm
\begin{tabular}{|l|*{6}{c}|c|}
\hline
external legs  	& $\phantom{i}$ 4 $\phantom{i}$ &  $\phantom{i}$ 5$\phantom{x}$&  $\phantom{i}$ 6$\phantom{x}$& $\phantom{x}$  7  $\phantom{x}$ &  $\phantom{x}$ 8 $\phantom{x}$ &  $n$ \\
\hline
amplitudes $A_{n;1}$    & 3 & 12 & 60 & 360 &  2520  & $\frac{1}{2}(n-1)!$ \\
KK-like relations              & 0 & 0 & 1 & 15 & 176 & $\frac{1}{2}(n-3)(n-2)!-c(n-1,3)$  \\
KK-indep. ampls.            & 3 & 12 & 59 & 345 &  2344 &  $c(n-1,3)+(n-2)!$\\
\hline
\end{tabular}
\vskip .5 cm
\end{table*}

As a first application of this principle, we work out the amplitude relations that only involve constant numerical coefficients~\eqref{KKlikeRel}. The result is displayed in Table~\ref{NPointTable2}. At four and five points we find no such KK-like relations. At six points there is one relation, and at seven and eight points there are 15 and 176 independent relations, respectively. 
Moreover, if we count the independent one-minus amplitudes remaining after taking into account the relations, we find $c(n-1,3)+(n-2)!$, where $c(n-1,3)$ are the same Stirling numbers that appear in Table~\ref{NPointTable} for the count of independent all-plus amplitudes. Similarly, $(n-2)!$ is the count of KK-independent amplitudes at tree-level. Indeed, the result of Table~\ref{NPointTable2} is a manifestation of the principle we introduced. The count of independent one-minus amplitudes under KK-like relations is the direct sum of the corresponding tree-level and one-loop all-plus counts.

What are the explicit relations then? Using the above principle we already know of one class of relations that should hold, these are the mentioned triple-photon decoupling relations~\eqref{tripid}. As discussed above, these relations hold for the all-plus amplitude and they also hold for tree amplitudes, since the latter satisfies single-photon decoupling relations (which imply that any number of photons decouple). For example, the single six-point relation in Table~\ref{NPointTable2} corresponds to eq.~\eqref{uni6pt}.

In the next section we will consider an additional class of particularly interesting relations that are satisfied simultaneously for
both tree-level and finite one-loop amplitudes. These will be in the form of eq.~\eqref{BCJlikeRel}.


\subsection{Additional relations for finite one-loop amplitudes} 

We will now show how to derive advanced relations based on the principle that one-minus relations are 
combination of all-plus and tree-level relations. Before we explicitly construct these new relations 
let us first give a schematic recipe.

The first step is to find an all-plus amplitude relation like one of those considered in section~\ref{allplusSection}. 
We will denote this expression, which vanishes upon evaluation, by $\mathcal{Q}^{\scriptsize \mbox{all-plus}}_n$, {\it i.e.}
\begin{align}
\mathcal{Q}^{\scriptsize \mbox{all-plus}}_n \,\equiv \,\sum_{\sigma} j_\sigma \, A_{n;1}^{\oneloop, \scriptsize \mbox{all-plus}}(\sigma) \, \mapsto \, 0\,,
\label{sch_plus}
\end{align}
for some non-vanishing integers $j_\sigma$. We then make the swap $A_{n,1}^{\oneloop, \scriptsize \mbox{all-plus}} \rightarrow A_n^{\mathrm{tree}}$, such that
\begin{align}
\mathcal{Q}^{\scriptsize \mbox{all-plus}}_n \quad \longrightarrow \quad \mathcal{Q}^{\mathrm{tree}}_n\,.
\end{align}
That is, we take the all-plus $A_{n,1}^\oneloop$ amplitudes in  eq.~\eqref{sch_plus} and substitute them with $A_n^{\mathrm{tree}}$ for some
arbitrarily helicity configuration. If it turns out that also $\mathcal{Q}^{\mathrm{tree}}_n = 0$, as is the case with eq.~\eqref{tripid},
we are done. According to our principle this combination of amplitudes should also vanish for the one-minus amplitudes. If $\mathcal{Q}^{\mathrm{tree}}_n$ evaluates to a nonzero expression, $\mathcal{Q}^{\mathrm{tree}}_n \neq 0$, then we need to find a tree-level relation involving $\mathcal{Q}^{\mathrm{tree}}_n$. For example, a linear combination of monomials $s_{ij}\mathcal{Q}^{\mathrm{tree}}_n$, where $s_{ij} \equiv (p_i+p_j)^2$, that vanishes due to tree-level BCJ relations. In general, finding such a relation seems like a difficult task since the $\mathcal{Q}^{\mathrm{tree}}_n$ are not regular tree amplitudes.  Nonetheless, with a clever choice of relation in the first step, in eq.~\eqref{sch_plus}, this actually becomes straightforward, as we will show below.

Assuming that such a linear combination is found, we have
\begin{align}
0 = \sum_{\sigma} P^{(1)}_{\sigma}(s_{ij}) \mathcal{Q}^{\mathrm{tree}}_n(\sigma)\, ,
\end{align}
which by construction is satisfied for both the all-plus one-loop amplitudes {\em and} the tree-level amplitudes. Invoking our principle, we then have a relation that will also be satisfied by the one-minus one-loop amplitudes. 

We will now look for examples of explicit relations of the kind eq.~\eqref{BCJlikeRel} with $d=1$.
The combination of amplitudes we will use for the first step in the construction is
\begin{align}
\mathcal{Q}_n(1,2,\ldots,n) \equiv 6A_n(1,2,\ldots,n)
- \sum_{k=2}^{n-1}\left[ \sum_{\sigma_k\in\mathrm{OP}(\{\alpha_k\}\cup \{\beta_k\})}
\hspace{-0.5cm} A_n(1,\{\sigma_k\}) \right]\,,
\label{Qn}
\end{align}
where the ordering of the arguments in $\mathcal{Q}_n$ is simply dictated by the ordering of the first amplitude on the right-hand side.
We have not yet specified which amplitudes the $A_n$'s stand for, (\textit{i.e.} whether tree-level, one-loop all-plus or one-loop one-minus).
When the $A_n$'s are one-loop all-plus amplitudes this is nothing but eq.~\eqref{allplus3}, so it follows $\mathcal{Q}^{\scriptsize \mbox{all-plus}}_n(1,2,\ldots,n)=0$.
The reason we choose this particular relation as our starting point is because of the following nice property. 
When we take the amplitudes in eq.~\eqref{Qn} to be tree-level amplitudes, the expression inside the brackets vanishes, 
and we are left with
\begin{align}
\mathcal{Q}^{\mathrm{tree}}_n(1,2,\ldots,n) = 6A_n^{\mathrm{tree}}(1,2,\ldots,n)\,.
\end{align}
With the help of eq.~\eqref{BCJ} it is now very easy to write down relations satisfied by $\mathcal{Q}^{\mathrm{tree}}_n$,
for instance
\begin{align}
0 ={}& s_{12}\mathcal{Q}^{\mathrm{tree}}_n(1,2,3,\ldots,n) + (s_{12}+s_{23})\mathcal{Q}^{\mathrm{tree}}_n(1,3,2,4,\ldots,n) \nonumber \\
&+ (s_{12}+s_{23}+s_{24})\mathcal{Q}^{\mathrm{tree}}_n(1,3,4,2,5,\ldots,n)  + \cdots  \nonumber \\
&+ (s_{12}+s_{23}+s_{24}+\cdots + s_{2(n-1)})
\mathcal{Q}^{\mathrm{tree}}_n(1,3,4,\ldots,n-1,2,n)\,,
\label{loopBCJ}
\end{align}
and of course all other independent BCJ relations as well.
{}From the properties of eq.~\eqref{Qn}, we have thus constructed $n$-point relations such as
\begin{align}
0 ={}& s_{12}\mathcal{Q}_n(1,2,3,\ldots,n) + (s_{12}+s_{23})\mathcal{Q}_n(1,3,2,4,\ldots,n) \nonumber \\
&+ (s_{12}+s_{23}+s_{24})\mathcal{Q}_n(1,3,4,2,5,\ldots,n)  + \cdots  \nonumber \\
&+ (s_{12}+s_{23}+s_{24}+\cdots + s_{2(n-1)})
\mathcal{Q}_n(1,3,4,\ldots,n-1,2,n)\,,
\label{genBCJ}
\end{align}
which are satisfied simultaneously by all-plus one-loop amplitudes (trivially, since each $\mathcal{Q}_n$ is zero here)
and tree-level amplitudes (on account of tree-level BCJ-relations).
We claim that all these relations are then also satisfied for one-minus one-loop amplitudes. We have explicitly
confirmed this up to $n=10$ using the full analytical expressions for the one-minus amplitudes.

\section{Further relations for finite amplitudes\label{moreSection}}
Given the types of relations for finite amplitudes that we have seen so far, it is natural to wonder if there exist additional relations
to explore. At tree level, the reduction of independent amplitudes to a basis of size $(n-3)!$, is explained naturally
from a string perspective
by the fact that we have to fix three world-sheet coordinates to define the string theory measure. This was shown
in~\cite{BDVmonodromy} to lead, in a simple manner, to both KK and BCJ relations for $n$ point tree level
amplitudes. At one-loop level the situation is naturally more complicated, but it would be interesting to
investigate if the basis of independent amplitudes also obeys the $(n-3)!$ counting.
Relations leading to such a reduction should be more complex from the point of view of monodromy, due to the 'internal' integration of the world-sheet loop.
In analogy with the tree-level basis reduction, entirely new structures with momentum-dependent coefficients may begin to appear.
We have done a partial investigation in this direction, which is briefly described in this section.

To start, let us again return to the simple four-point level. Using the explicit expression for the known analytic formula for the
$(-++\,+)$ amplitude \eqref{oneMinusAmp}, we have
\begin{align}
0 ~=~ A_{4;1}^\oneloop(1^-,2^+,3^+,4^+)s_{14}^3 -A_{4;1}^\oneloop(1^-,2^+,4^+,3^+) s_{13}^3\,.
\label{cubicRel}
\end{align}
This relation is reminiscent of the corresponding four-point tree-level BCJ relation, but instead of a single power of $s_{ij}$ it involves
{\em cubes} of $s_{ij}$'s. This is a new and unusual feature that begs for an interpretation, perhaps in terms of monodromy in string theory. Using eq.~\eqref{cubicRel} we may eliminate all but one of the four-point amplitudes. 

More complicated structures appear for higher point amplitudes. At five-point for the all-plus amplitude we have four independent quadratic relations, such as 
\def\A(\{#1,#2,#3,#4,#5\}){A_{5;1}^\oneloop(#1^+\!,#2^+\!,#3^+\!,#4^+\!,#5^+)}
\def\s(#1,#2){s_{#1#2}}
\begin{equation}\begin{split}0&=
\A(\{1,4,2,3,5\})\left(\s(1,5)^2\!+\!4 \s(4,5) \s(1,5)\!+\!\s(1,2) (\!-\!5 \s(1,5)\!+\!4 \s(2,3)\!-\!2 \s(3,4))\!-\!2 \left(\s(2,3)^2\!+\!\s(3,4) \s(2,3)\!-\!\s(3,4)^2\right)\right)
\\&
+\A(\{1,3,2,4,5\})\left(\!-\!\s(1,5)^2\!+\!4 \s(4,5) \s(1,5)\!+\!\s(1,2) (\!-\!3   \s(1,5)\!+\!4 \s(2,3)\!-\!2 \s(3,4))\!+\!2 \left(\s(2,3)^2\!+\!\s(3,4) \s(2,3)\!+\!\s(3,4)^2\right)\right)
\\&
+\A(\{1,4,3,2,5\})\left(3 \s(1,5)^2\!+\!4 \s(4,5) \s(1,5)\!+\!4 \s(4,5)^2\!-\!2 \left(\s(2,3)^2\!-\!\s(3,4)
   \s(2,3)\!+\!\s(3,4)^2\right)\!+\!\s(1,2) (\s(1,5)\!+\!2 \s(3,4)\!-\!4 \s(4,5))\right)\\&
+\A(\{1,2,3,4,5\})\left((\s(1,5)\!+\!2 \s(4,5))^2\!-\!2 \left(\s(2,3)^2\!+\!\s(3,4) \s(2,3)\!+\!\s(3,4)^2\right)
\!+\!\s(1,2) (3 \s(1,5)\!-\!4 \s(2,3)\!+\!2
   \s(3,4)\!-\!4 \s(4,5))\right) \\&
+\A(\{1,2,4,3,5\})\left(\!-\!(\s(1,5)\!+\!2 \s(4,5))^2\!+\!2 \left(\s(2,3)^2\!+\!\s(3,4) \s(2,3)\!+\!\s(3,4)^2\right)\!+\!\s(1,2) (5 \s(1,5)\!-\!4 \s(2,3)\!-\!2 \s(3,4)\!+\!4 \s(4,5))\right)
 \\&
+\A(\{1,3,4,2,5\})\left(\s(1,5)^2\!-\!4 \s(4,5) \s(1,5)\!+\!2 \s(2,3)^2\!-\!4 \s(4,5)^2\!-\!2 \s(3,4) (\s(2,3)\!+\!\s(3,4))\!+\!\s(1,2) (3 \s(1,5)\!+\!2 \s(3,4)\!+\!4 \s(4,5))\right)\,,
 \end{split}  \end{equation}
effectively reducing the number of independent amplitudes to only two. 

For the one-minus amplitude at five points we find six independent relations with kinematic invariants of
power four and five, {\it e.g.} we have  
\def\B(\{#1,#2,#3,#4,#5\}){A_{5;1}^\oneloop(#1^-\!\!,#2^+\!\!,#3^+\!\!,#4^+\!\!,#5^+)}
\def\s(#1,#2){s_{#1#2}}
\begin{equation}\begin{split}
0&=\B(\{1,2,3,4,5\}) \s(1,2)^4\!+\!\B(\{1,3,2,4,5\}) \s(1,2)^4\!
+\!\B(\{1,2,4,3,5\}) \Big(\s(1,5) \s(2,3)^3\!-\!(\s(1,2)\!+\!2 \s(1,5)\!\\&+\!8 \s(3,4)) \s(4,5)^3\!+\!(4 \s(1,5)^2\!-\!(\s(2,3)\!-\!10 \s(3,4)) \s(1,5)\!+\!(8 \s(2,3)\!-\!17 \s(3,4))
   \s(3,4)) \s(4,5)^2\!+\!(3 \s(1,5) \s(2,3)^2\!-\!\s(3,4)^3) \s(4,5)\Big)\!
\\&+
\B(\{1,2,5,3,4\}) c_1 \!
+
\B(\{1,3,4,2,5\})c_2
+\B(\{1,4,3,2,5\}) c_3\\&
+\B(\{1,4,2,3,5\}) c_4
+\B(\{1,2,3,5,4\}) c_5
+\B(\{1,2,4,5,3\}) c_6\\&
+\B(\{1,3,2,5,4\}) c_7
+\B(\{1,2,5,4,3\}) c_8
+\B(\{1,3,5,2,4\}) c_9\,,
   \end{split} \end{equation}
where $c_i$ are complicated degree-four polynomials in the kinematic invariants ($s_{12}$, $ s_{23}$, $s_{34}$, $s_{45}$, $s_{15}$).

In these expressions we have different combinations of various products of Mandelstam invariants $s_{ij}$ to the fourth and fifth power respectively. The totality of such relations together with the previously discussed linear relations surprisingly reduces the number of independent amplitudes at five point to a total of two independent amplitudes. This again matches the $(n-3)!$ reduction that has been observed at tree level.

At six point we have found additional relations between amplitudes. For the all-plus case there exists a set of relations reducing the number of independent amplitudes to just three. This yields a reduction that is better than the expected $(n-3)!$ counting.  It should be kept in mind, however, that the relations that we discuss here are specific to the given helicities. Even at tree level there exists reductions beyond the $(n-3)!$ counting for specific helicity sectors. As an illustrative example, we have considered all relations for six-point MHV amplitudes at tree level. These reduces the number of independent amplitudes from the universal six to just two.  An example of such a relation is given by
\def\R(\{#1,#2,#3,#4,#5,#6\}){A_6^{\rm tree}(#1^-\!\!,#2^-\!\!,#3^+\!\!,#4^+\!\!,#5^+\!\!,#6^+)}
\def\s(#1,#2){s_{#1#2}}
\begin{equation}\begin{split}
0&=
   \R(\{1,2,4,6,3,5\}) (\s(1,3)\!+\!\s(3,5)) (\s(1,2)\!+\!\s(1,3)\!-\!\s(1,6)\!+\!2 \s(2,3)\!+\!\s(3,4)\!+\!\s(3,5))\!\\&
+\!\R(\{1,2,6,4,3,5\}) (\s(1,3)\!+\!\s(3,5)) (\s(1,2)\!+\!\s(1,3)\!-\!\s(1,6)\!+\!2 \s(2,3)\!+\!2 \s(2,4)\!+\!\s(3,4)\!+\!\s(3,5))\!\\&
+\!\R(\{1,2,3,5,4,6\}) \s(1,6)(\s(2,4)\!-\!\s(3,4)\!+\!\s(4,5))\!\\&
+\!\R(\{1,2,3,5,6,4\}) (\s(1,6)\!-\!2 (\s(2,3)\!+\!\s(3,4))) (\s(1,2)\!+\!\s(1,3)\!+\!\s(2,3)\!+\!\s(2,4)\!+\!\s(3,4)\!-\!\s(5,6))\!\\&
+\!\R(\{1,2,4,5,6,3\}) (\s(1,6)\!-\!\s(2,3)\!-\!\s(3,4)\!-\!\s(3,5))(\s(1,2)\!+\!\s(2,3)\!+\!\s(2,4)\!+\!\s(5,6))\!\\&
-\!\R(\{1,2,4,6,5,3\}) (\!-\!\s(1,6)\!+\!\s(2,3)\!+\!\s(2,4)\!+\!\s(3,4)\!+\!\s(3,5)\!-\!\s(5,6)) (\s(1,2)\!+\!\s(2,3)\!+\!\s(2,4)\!+\!\s(5,6))\!\\&
-\!\R(\{1,2,6,4,5,3\}) (\!-\!\s(1,6)\!+\!\s(2,3)\!+\!\s(2,4)\!+\!\s(3,4)\!+\!\s(3,5)\!-\!\s(5,6))(\s(1,2)\!+\!\s(2,3)\!+\!\s(2,4)\!+\!\s(5,6))\!\\&
-\!\R(\{1,2,4,3,5,6\}) \s(1,6) (\s(1,2)\!+\!\s(1,3)\!+\!\s(2,3)\!+\!\s(3,5)\!+\!\s(5,6))\!\\&
-\!\R(\{1,2,4,3,6,5\}) (\s(1,6)\!+\!\s(5,6)) (\s(1,2)\!+\!\s(1,3)\!+\!\s(2,3)\!+\!\s(3,5)\!+\!\s(5,6))\!\\&
+\!\R(\{1,2,6,3,5,4\}) (\s(1,2)\!+\!\s(1,3)\!+\!\s(2,3)\!+\!\s(2,4)\!+\!\s(3,4)\!-\!\s(5,6)) (\s(1,2)\!+\!\s(1,3)\!+\!\s(2,3)\!+\!\s(2,4)\!+\!\s(3,5)\!+\!\s(5,6))\!\\&
+\!\R(\{1,2,6,3,4,5\}) (\s(1,3)\!-\!\s(1,6)\!+\!\s(2,3)\!+\!\s(2,4)\!+\!2 \s(3,4)\!+\!\s(3,5)\!-\!\s(5,6))(\s(1,2)\!+\!\s(1,3)\!+\!\s(2,3)\!+\!\s(2,4)\!+\!\s(3,5)\!+\!\s(5,6))\!\\&
+\!\R(\{1,2,3,6,5,4\}) (\s(1,2)\!+\!\s(1,3)\!+\!\s(2,3)\!+\!\s(2,4)\!+\!\s(3,4)\!-\!\s(5,6)) (\s(1,2)\!+\!\s(1,6)\!-\!\s(2,3)\!+\!\s(2,4)\!-\!2 \s(3,4)\!+\!2 \s(5,6))\!\\&
-\!\R(\{1,2,3,4,5,6\}) \s(1,6) (2\s(1,2)\!+\!\s(1,3)\!+\!2 \s(2,3)\!+\!\s(2,4)\!+\!\s(3,4)\!+\!\s(3,5)\!+\!2 \s(5,6))\!\\&
+\!\R(\{1,2,3,4,6,5\}) \big(\s(2,3)^2\!-\!2 \s(1,6) \s(2,3)\!+\!2 \s(2,4)\s(2,3)\!+\!\s(3,4) \s(2,3)\!+\!\s(2,4)^2\!-\!2 \s(5,6)^2\!-\!\s(1,3) \s(1,6)\!\\&-\!\s(1,6) \s(2,4)\!-\!\s(1,6)
   \s(3,4)\!+\!\s(2,4) \s(3,4)\!-\!\s(1,6) \s(3,5)\!+\!\s(1,2) (\!-\!2 \s(1,6)\!+
\!\s(2,3)\!+\!\s(2,4)\!+\!\s(3,4)\!-\!2 \s(5,6))\!-\!(\s(1,3)\!+\!2 \s(1,6)\!+\!\s(2,3)\!+\!\s(3,5)) \s(5,6)\big)\!\\&
+\!\R(\{1,2,6,5,4,3\}) \big(\s(5,6)^2\!-\!(\s(1,2)\!+\!\s(2,3)\!+\!\s(2,4))^2\big)\!\\&
+\!\R(\{1,2,3,6,4,5\}) \big(\!-\!\s(1,2)^2\!-\!(2 \s(1,3)\!+\!\s(1,6)\!+\!3 \s(2,3)\!+\!\s(3,4)\!+\!\s(3,5)\!-
\!\s(4,5)\!+\!\s(5,6)) \s(1,2)\!-\!\s(1,3)^2\!-\!2 \s(2,3)^2\!+\!\s(2,4)^2\!\\&-\!2 \s(3,4)^2\!-\!2\s(5,6)^2\!-\!\s(1,6) \s(2,3)\!-\!\s(1,6) \s(2,4)\!-\!\s(2,3) \s(2,4)\!+\!\s(1,6) \s(3,4)\!-\!3 \s(2,3) \s(3,4)\!-\!\s(2,4) \s(3,4)\!-\!\s(1,6) \s(3,5)\!-\!\s(2,3) \s(3,5)\!\\&-\!\s(1,3) (4 \s(2,3)\!+\!\s(2,4)\!+\!2 \s(3,4)\!+\!\s(3,5)\!-\!\s(4,5))\!-\!\s(1,6) \s(4,5)\!+\!3 \s(2,3)
   \s(4,5)\!+\!\s(2,4) \s(4,5)\!+\!2 \s(3,4) \s(4,5)\!+\!\s(3,5) \s(4,5)\!\\&+\!(\!-\!3 \s(1,6)\!+\!2 \s(2,3)\!+\!\s(2,4)\!+\!4 \s(3,4)) \s(5,6)\big)\,.
\end{split}\end{equation}
This relation does not hold for six-point NMHV configurations. At higher points there are similar particular relations that do not hold for general, only specific, helicities.  In this sense there is a fundamental issue on how to define a basis set of amplitudes at one loop, using only the information from the finite amplitudes, which are of a particular helicity type. Nonetheless, it is interesting to observe that the patterns of identities we have enlightened in this section are very reminiscent of monodromy relations. And for the all-plus and one-minus finite loop amplitudes we have found no solid evidence that goes against a universal $(n-3)!$ reduction for the basis of independent amplitudes.

\section{Conclusions\label{conclusion}}
In this paper we have demonstrated that the finite one-loop gauge theory amplitudes have a rich structure of amplitude relations, which much resembles that of tree-level Kleiss-Kuijf and BCJ amplitude relations, or equivalently, monodromy relations. By investigating two types of finite amplitudes, the pure-gluon amplitudes  with helicity $(+++\cdots+)$ and $(-++\cdots+)$, we have found interesting patterns, some of which may very well generalize to other finite contributions at one loop. The all-plus-helicity amplitudes satisfy relations that involve simple sums and differences of different color-ordered amplitudes. As we have shown, these KK-like relations can be entirely understood using a diagrammatic formalism built on kinematical graphs with exactly one symmetric quartic vertex, and an arbitrary number of antisymmetric cubic vertices. For the one-minus-helicity amplitudes, similar amplitude relations can be explained by adding purely cubic diagrams to the previous types of diagrams. This leads directly to triple-photon decoupling relations, which holds for both types of finite one-loop amplitude considered here.

One interesting pattern that emerges is that amplitude relations that are valid at both tree level and for the one-loop all-plus case, also always seem to hold for the one-minus case. This provides a useful principle for how to construct amplitude relations for the more complex one-minus amplitudes.  Using precisely this principle we have obtained new relations that involve sums of amplitudes multiplied by kinematical invariants. 

We have also found relations that appear to not fit any simple pattern expected from tree level. For example, a curious four-point relation involving cubic monomials appears in the one-minus case. Interestingly, one can use this relation to eliminate all but one of the four-point amplitudes, similar to BCJ relations at tree level. Also at five and six points there are intricate relations that allows one to eliminate all but $(n-3)!$ amplitudes for both types of finite amplitudes, in analogy with tree level. However, at six points there appears to be even stronger reductions possible, which we explain by the
detailed relations being specific to the given helicity sector. We have shown how similar non-universal relations exist at tree level for the 
MHV amplitudes, which support this interpretation. 

It would be of clear interest to understand the observed relations better. For example, there are strong hints that monodromy in string theory can provide a clarification of the patterns.
Similarly, introducing kinematic Jacobi relations for the diagrams used in this paper, would likely explain some of the observed relations.
It should also be worthwhile to explore deeper the relations using other field-theoretic methods. 

An interesting question is, obviously, if similar structures can be deduced for loop amplitudes that have divergences. 
The closest analogue
of the finite amplitudes would be the rational pieces of more general one-loop amplitudes, which is an important subject in current next-to-leading-order calculations of cross sections (see {\it e.g.}~\cite{bootstrapping}). For these amplitudes the analysis is expected to be more complicated due to the problem of separating finite and divergent pieces in a consistent manner. We leave this question to future work. 

\section*{Acknowledgements}
We thank Zvi Bern, John Joseph Carrasco and Pierre Vanhove for discussions. H.J. is grateful for the hospitality of the Niels Bohr International Academy, where part of this work was carried out.
This work was supported by the European Research Council under Advanced Investigator Grant ERC-AdG-228301.

\end{document}